\documentstyle[aps,prc,epsfig,amsfonts,mymcite,floats]{revtex}

\begin{document}

\wideabs{
\title {On the equivalence of pairing correlations and intrinsic vortical
  currents\\ in rotating nuclei}

\author{H. Laftchiev$^{1,2}$, D. Sams\oe{}n$^1$ P. Quentin$^{1,3}$, and
  I. N. Mikhailov$^{4,5}$}

\address{$^1$Centre d'\'Etudes Nucl\'eaires de Bordeaux-Gradignan
  (CNRS-IN2P3 and Universit\'e Bordeaux I), Le Haut Vigneau BP 120, F-33175 
  Gradignan, France \\
$^2$Institute of Nuclear Research and Nuclear Energy (Bulgarian Academy of
  Sciences), Tzarigradsko Chaussee 72, 1784 Sofia, Bulgaria \\
$^3$Theoretical Division, T-DO (Los Alamos National Laboratory), PO Box 1663,
  Los Alamos, NM 87545, USA \\
$^4$Bogoliubov Laboratory of Theoretical Physics (Joint Institute of Nuclear
  Research), Joliot-Curie str. 6, 141980 Dubna (Moscow Region), Russia \\
$^5$Centre de spectrom\'etrie Nucl\'eaire et de Spectrom\'etrie de Masse, 
  (CNRS-IN2P3 and Universit\'e Paris XI), B\^at. 104, F-91406 Orsay-Campus, 
  France}
\maketitle

\begin{abstract}
The present paper establishes a link between pairing correlations in rotating
nuclei and collective vortical modes in the intrinsic frame.
We show that the latter can be embodied by a simple S-type coupling {\it \`a
la} Chandrasekhar between rotational and intrinsic vortical collective modes.
This results from a comparison between the solutions of microscopic
calculations within the HFB and the HF Routhian formalisms.
The HF Routhian solutions are constrained to have the same Kelvin circulation
expectation value as the HFB ones.
It is shown in several mass regions, pairing regimes, and for various spin
values that this procedure yields moments of inertia, angular velocities, and
current distributions which are very similar within both formalisms.
We finally present perspectives for further studies.
\end{abstract}
}

\narrowtext

\section{Introduction and methods}
\label{sec:intro}
In a recent paper \cite{SQM99:period}, some of the authors of the present
article conjectured that the well-known dynamical effects of pairing
correlations in rotating nuclei, explicited in the significant decrease of the
moments of inertia from rigid body values \cite{BM55,Bel61}, could be pictured
as a vortical intrinsic collective motion coupled to the global rotation of
the so-called intrinsic (body-fixed) frame relatively to the laboratory
system.  Namely, one intuitively would think of such a mode as aligned and
counter-rotating with respect to the global angular velocity so as to produce
the above quoted reduction of the moments of inertia.  Furthermore, if one
wants to preserve the nuclear shape in the presence of this intrinsic vortical
mode the latter should be tangential to the grossly-defined nuclear surface.

As often suggested after R. Y. Cusson \cite{Cus68} (see also
Ref. \cite{MQS97:vort} for some relevant references), a simple ansatz for such
a coupling is provided by the S-ellipsoid fluid dynamics as studied
classically in great details by Chandrasekhar in the context of celestial
self-gravitating objects \cite{Chandra}.  It makes use of a velocity field
which is linear in the coordinates and produces a coupling of a global
rotation with an intrinsic mode corresponding to a motion tangential to the
body surface which is supposed to be ellipsoidal.
Furthermore, the vorticities (i.e., the curl) of the two velocity fields are
assumed to be aligned or anti-aligned along a principal axis of the ellipsoid.
As shown, e.g., in Ref. \cite{MBQ96} (see Fig.~1 there) such a simple
parameterization is capable of yielding a variety of modes in the laboratory
frame from rigid-body rotation to irrotational modes as well as various shear
modes.

The connection of this S-ellipsoid fluid dynamics with pairing correlations may
be discussed at various levels. 
First, one may note that a tangential intrinsic mode which is not changing the
body shape (in the $\vec r$-sector of the phase-space) is only able to
redistribute the density function in the $\vec p$-sector of the phase-space.
Clearly, up to self-consistent couplings between the momentum redistribution
and the body shape, the BCS pairing correlations behave exactly in the same
way (redistributing the single-particle momenta).
Second, in the semiclassical (Thomas-Fermi) calculations of Ref. \cite{DSK85},
upon widely varying the pairing strength in some deformed rare-earth rotating
nuclei, the authors have unambiguously exhibited current patterns very similar
indeed to the classical current patterns of S-ellipsoids.
Such currents had been shown before to exist in paired solutions at finite
spin \cite{KM79,FKM80}. They were however embedded in the midst of
shell-effect---generated intrinsic currents whose existence had been found
long ago (see the non-paired rotating solutions of \cite{Rad76}) and which
will be further discussed in Section \ref{ssec:gado} below.

However, the arguments above sketched are purely qualitative. It is the aim
of this paper to investigate how quantitatively correct they might be.

To achieve this goal, it is first necessary to find a tool to quantify the
intrinsic vortical content of a given current distribution. Within the context
of S-ellipsoid fluids, it is rather natural to use the Kelvin circulation
whose components are defined by (see, e.g., \cite{Ros92})
\begin{equation}
  \hat K_k = \sum_{i,j} \epsilon_{ijk}\left(\frac{a_j}{a_i}x_ip_j -
    \frac{a_i}{a_j}x_jp_i\right)
\end{equation}
where $\epsilon_{ijk}$ is the totally antisymmetrical third rank tensor
and the $a_i$'s are characteristic lengths associated with the ellipsoidal
shape of the nuclei (e.g., the semi-axes).
These lengths are proportional to $\sqrt{\langle x_i^2\rangle}$.
These operators clearly appear as the components of a doubly-stretched (in
both $\vec r$ and $\vec p$) orbital angular momentum. Physically, they
correspond to a rotation in the intrinsic frame of a sphere obtained by
stretching the ellipsoidal intrinsic distribution. Indeed as shown by Lebowitz
(see, e.g., Ref. \cite{Chandra}), the S-ellipsoid coupling mode may be
understood as a global rotation along some principal axis $\vec u$, followed
by the stretching to a sphere and the rotation (according to the same axis
$\vec u$ of the considered ellipsoid) as above described, and finally by the
inverse stretching back to the original ellipsoidal shape.

To check the validity of the above mentioned conjecture, we compare the
results of two types of microscopic calculations, one involving pairing
correlations in a purely rotating formalism, the other with no pairing but
imposing the S-ellipsoid dynamics for the global rotation and intrinsic
vortical mode velocity fields as well as for their coupling.

The pairing correlations are treated {\it \`{a} la} Hartree-Fock-Bogoliubov
(HFB) using a Skyrme effective interaction in the particle-hole channel and a
seniority force in the pairing channel within a code developed by H. Laftchiev
\cite{Laf00:th,Laf01:nobel}. The triaxial character of
the solutions arising from the time-reversal symmetry breaking inherent to
rotating nuclei is treated within the Fourier decomposition formalism
described in Ref. \cite{SQB99:HF}. Namely, the wavefunctions and relevant
densities are defined in cylindrical coordinates on a mesh in the two spatial
directions $(r,z)$ and as Fourier series in the azimuthal angle $\theta$. An
approximate description of particle-number projection within the Lipkin-Nogami
(LN) scheme \cite{Lip64a,*Lip64b,*Lip65,*Lip66,*Lip73} is also available in the
code.

The S-ellipsoid coupling case is described within the Hartree-Fock-Skyrme
generalized Routhian formalism of Refs. \cite{SQB99:HF,Generouth}. This
formalism makes it possible to describe the S-ellipsoid dynamics upon using a
double constraint on both angular momentum and Kelvin circulation via two
angular velocities $\Omega$ and $\omega$ (relevant to each collective mode)
along the first-axis.

These two microscopic approaches have been used in the following scheme.
In a first step, we have used the HFB formalism with a constraint on angular
momentum using realistic pairing strengths. We thus
generate vortical currents in the intrinsic frame which can be quantified for
instance by the mean value of the Kelvin circulation operator measured in the
intrinsic frame and defined as
\begin{equation}
  \label{eq:kelvin}
  \langle\hat K_1\rangle = \int (\tilde{\vec r} \times
  \tilde{\vec j})_1\; {\rm d}^3r \,.
\end{equation}
There $\tilde{\vec r}$ is the stretched position vector and $\tilde{\vec j}$
is the stretched current in the intrinsic frame whose components are obtained
after subtracting the rotating-part to the usual current-density $\vec j$ as
\begin{equation}
  \tilde j_i = a_i \left[\vec j - \frac{m}{\hbar} (\vec\Omega\times\vec r)
    \rho\right]_i \,.
\end{equation}
where $\vec j$ is the usual quantum mechanical current density defined as
\begin{equation}
  \label{eq:current}
  \vec j = \frac{1}{2i} (\vec\nabla_{\vec r\,'}-\vec\nabla_{\vec r}) 
 \rho(\vec r,\vec r\,')\bigg|_{\vec r=\vec r\,'}
\end{equation}
The mean value $I$ of the angular momentum operator is defined through
\begin{equation}
  \label{eq:angular}
  \sqrt{I(I+1)} = \langle\hat I\rangle = \int (\vec r \times \vec j + \vec\rho
  )_1\; {\rm d}^3r
\end{equation}
where $\vec\rho$ is the spin-vector density.

In a second step, we perform pure Hartree-Fock calculations (i.e., without
pairing) with a double constraint on both angular momentum and Kelvin
circulation to obtain the same values $I$ and $\langle\hat K_1\rangle$ as in
the preceding step.  In the following, this type of calculations will be
referred to as HF+V calculations.

The paper will be organized as follows. In the next Section we will present
and compare the results of the numerical calculations yielded by the two
formalisms, namely HFB and HF+V in several mass regions. Section
\ref{sec:conclu} will be devoted to a summary of the salient features
exhibited in Section \ref{sec:results} and to some
perspectives for further studies.

\begin{table}[t]
  \caption{Pairing strength $G_n$ and $G_p$ for neutrons and protons as well
    as energy cutoff $\mu$ (in MeV) for each of the three studied nuclei. The
    matrix elements $g_q$ of the seniority force are defined by the usual
    prescription $g_q=G_q/(11+N_q)$. The configuration space includes all
    single-particle states whose energies lies below $\epsilon+\mu$ where
    $\epsilon$ is the chemical potential.}
\vskip\baselineskip
    \begin{tabular}{lccc}
            & $^{150}$Gd & $^{192}$Hg & $^{254}$No\\ \hline
      $G_n$ & 14.4 & 11.5 & 13.8 \\
      $G_p$ & 14.4 & 11.5 & 15.0  \\ 
      Cutoff ($\mu$)&  5   & 10   &  7   
    \end{tabular}
    \label{tab:forces}
\end{table}

\section{Results}
\label{sec:results}
Numerical calculations have been performed for three heavy nuclei: two in the
$A\sim 150$ and $A\sim 190$ superdeformation regions (namely for the $^{150}$Gd
and the $^{192}$Hg nuclei) including LN approximate particle-number projection
scheme and one very heavy nucleus ($^{254}$No) without LN projection.  The
SkM* parameterization \cite{Bar82:SkM} of the Skyrme effective force has been
used. The pairing strengths $G_q$ for the charge state $q$ are given for each
calculated nuclei in table \ref{tab:forces} together with the corresponding
single-particle configuration space in use.  For all three nuclei, kinetic
(${\frak J}^{(1)}$) and dynamical (${\frak J}^{(2)}$) moments of inertia are
calculated using the formulae

\begin{figure*}[t]
  \begin{center}
  \epsfig{width=17cm,file=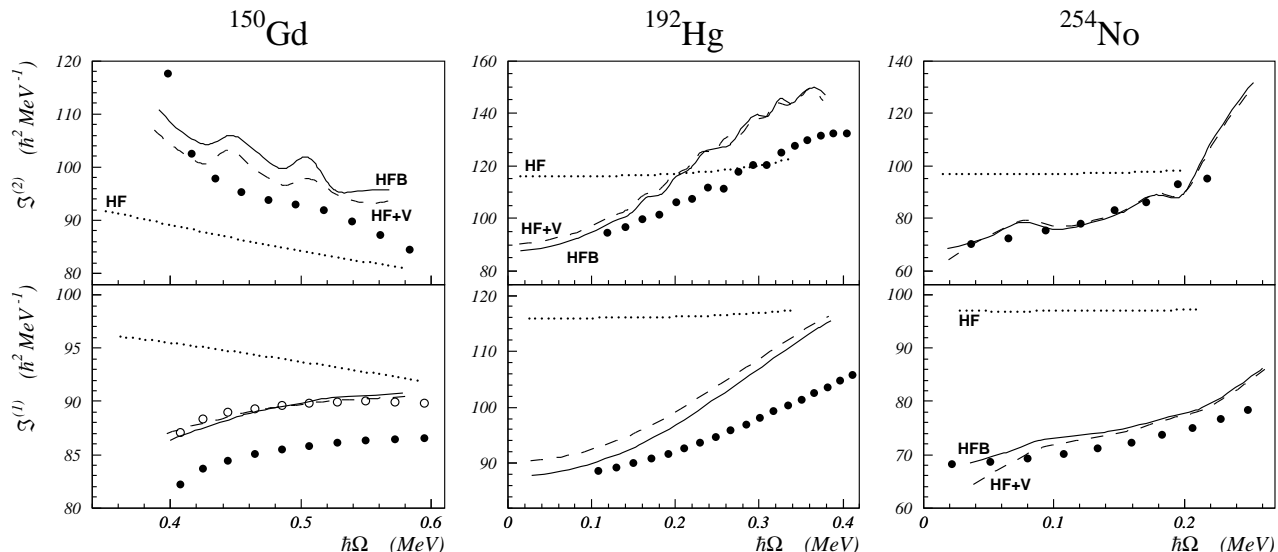}
  \caption{Dynamical (upper panel) and kinetic (lower panel) moments of
  inertia (in $\hbar^2\,$MeV$^{\text{-1}}$) for the three considered nuclei
  as functions of the angular velocity $\Omega$ in MeV.  The conventions in
  use are the following: HF value (dotted line), HFB value (full line), and
  HF+V value (dashed line). Experimental data are represented as full circles,
  excepted for the $^{150}$Gd kinetic moment of inertia with the assumption
  $I_{\mathrm{init}}=34\;\hbar$ as opened circles (lower left panel). The
  Lipkin-Nogami correction has been applied for $^{150}$Gd and $^{192}$Hg.}
  \label{fig:inertia}
  \end{center}
\end{figure*}

\begin{equation}
  \label{eq:kinetic}
  {\frak J}^{(1)} = \frac{\langle\hat I\rangle}{\Omega}\,,
\end{equation}
and
\begin{equation}
  \label{eq:dynamic}
  {\frak J}^{(2)} = \frac{\partial \langle\hat I\rangle}{\partial\Omega}\,.
\end{equation}
As was already mentioned in a previous paper \cite{Laf01:nobel}, using these
expressions for the moments of inertia, we assume that they can be computed as
well from the energy-corrected LN wavefunctions. We also assume that the
angular velocity dependence of the estimate of LN $\lambda_2$ parameter is
small, otherwise ${\frak J}^{(2)}$ would have to be evaluated through second
derivatives of the energy with respect to $\Omega$.

\subsection{The yrast superdeformed band of $^{150}$Gd }
\label{ssec:gado}
Up to now, fourteen superdeformed bands have been experimentally found in
$^{150}$Gd \cite{Fall91:gd,Erturk}. We will deal here with the yrast
superdeformed band only. Using the Cranked-HF approach, it was impossible
\cite{SQB99:HF} to reproduce the trends of the experimental kinetic and
dynamical moments of inertia of this band as can be seen in
Fig.~\ref{fig:inertia} (dotted line).

The results presented in the upper panel of Fig.~\ref{fig:inertia} for the
dynamical moment of inertia of this nucleus show a much better agreement with
experimental data within the HFB+LN formalism as compared to the HF
one. Indeed, we now reproduce the decrease of this moment of inertia along the
band with theoretical values which are rather close to the experimental ones
as was previously shown within pure HFB calculations by Bonche {\it et al.}
\cite{BFH96:gado}.
In the lower panel, the agreement of HFB+LN kinetic moment of inertia with the
experimental data of Ref. \cite{Fall91:gd}, though much better than the HF one,
is seen to be rather poor. However, the experimental value of this moment of
inertia relies on an assumption of the lowest spin value of the observed band
[see eq. (\ref{eq:kinetic})] which has been assumed in \cite{Fall91:gd} to be
$I_{\mathrm{init}}=32\;\hbar$.
Taking now for the initial spin of the band $I_{\mathrm{init}}=34\;\hbar$
shifts up the experimental data and then yields a very good agreement with
HFB+LN values of the moment of inertia.

\begin{figure*}[t]
  \begin{center}
  \epsfig{width=17cm,file=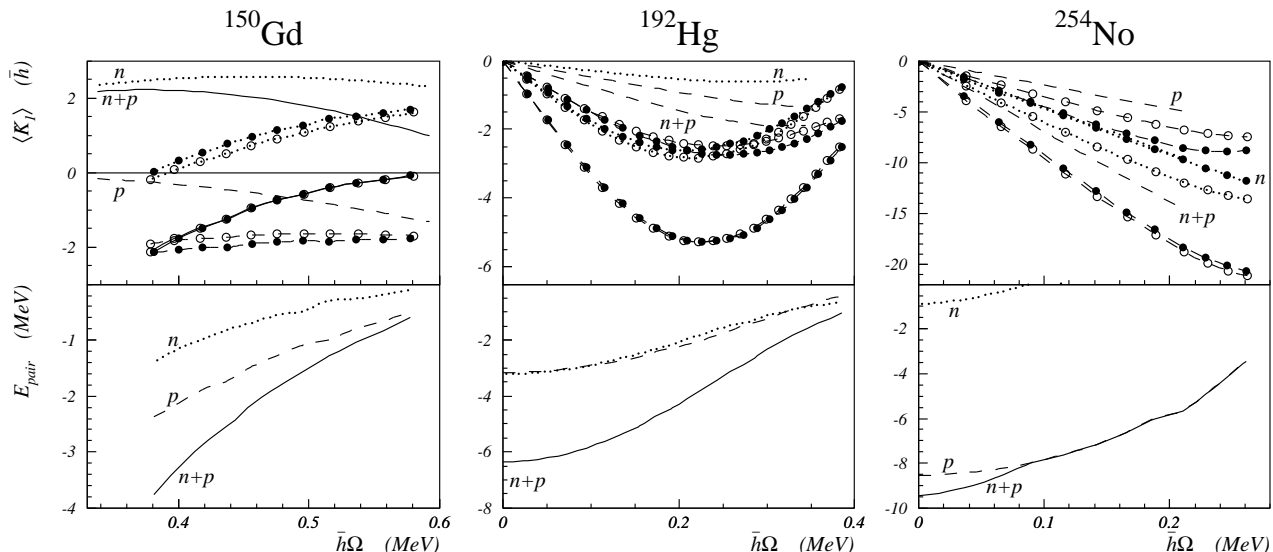}
  \caption{Kelvin circulation mean value in $\hbar$ units (upper panel) and
  pairing energy in MeV (lower panel) for the three considered nuclei as
  functions of the angular velocity $\Omega$ in MeV.  The conventions in use
  are the following: proton contributions (dashed line), neutron contributions
  (dotted line), and total contributions (full line); in the upper panel: HF
  value (no symbols), HF+V value (opened circles), HFB value (full
  circles). The Lipkin-Nogami correction has been applied for $^{150}$Gd and
  $^{192}$Hg.}
  \label{fig:kelvin}
  \end{center}
\end{figure*}

Let us now focus on the intrinsic vortical content of the current density. In
the upper part of Fig.~\ref{fig:kelvin} the Kelvin circulation mean value is
plotted as a function of the angular velocity for $^{150}$Gd.
 One may first notice that the Kelvin circulation mean value in the
HF case is not zero, even though the intrinsic vorticity constraint is
absent. It is negative (with a rather small absolute value of about 1~$\hbar$)
for the protons while it is positive and larger (about 2~$\hbar$) for the
neutrons. These contributions should most likely be attributed to small
inhomogeneities in the density due to shell effects. It is almost constant for
the neutrons and slightly decreasing with $\Omega$ for the protons.
Dubbing this current patterns as shell effects may also be understood in rather
general terms. In Ref. \cite{BBQ94} within purely semiclassical microscopic
calculations, it has been shown that for finite systems small surface-peaked
counter-rotating intrinsic currents do appear. This effect is analogous to the
diamagnetism which is observed in an electronic system submitted to a constant
magnetic field. Thence the origin of all intrinsic currents (in Routhian HF
calculations) which are not surface-peaked and counter-rotating (as obtained
years ago in, e.g., \cite{Rad76}) are clearly and properly to be attributed to
shell effects.

Comparing now the HF curves with the HFB ones, we see that the introduction of
pairing correlations shifts down the Kelvin circulation mean value by about
2~$\hbar$ for each charge state at low angular velocities and still
0.5 $\hbar$ units at the end of the band. This decrease of the Kelvin
circulation in the HFB case is related, as it has been already demonstrated in
\cite{SQM99:period}, to the fact that the collective effect of pairing
correlations is equivalent to a counter-rotating intrinsic vortical motion
(see also Ref. \cite{DSK85}).
The combination of the above mentioned HF variation of $\langle\hat
K_1\rangle$ with $\Omega$ for protons and neutrons together with the regular
decrease of the pairing correction as described previously yields an increase
of the neutron contribution and an almost constant pattern for the protons. As
a result the total Kelvin circulation exhibits an $\Omega$ variation similar
to the neutron one.

This decreasing pairing effect on the $^{150}$Gd Kelvin circulation is nicely
correlated, as can be seen on the lower panel of Fig.~\ref{fig:kelvin}, with
the decrease of the pairing energy (defined as a trace of the product of the
abnormal density with the gap potential).
The latter is a well-known behavior, known as the Mottelson-Valatin effect
\cite{MV60}, similar to the effect of a magnetic field on a superconductor
below the critical point \cite{Kittel:solids}.  Indeed as the pairing energy
almost vanishes at high angular velocity, the Kelvin circulation in the HFB
formalism almost reaches its HF value.

\begin{figure*}[ht]
  \begin{center}
  \epsfig{width=17cm,file=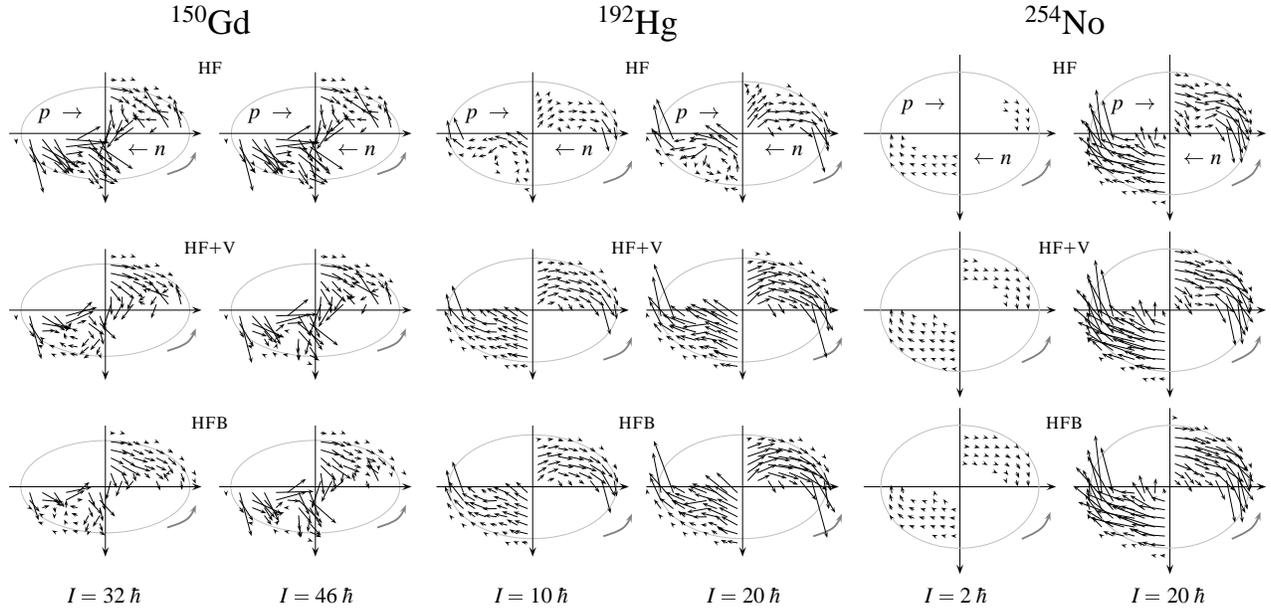}
  \caption{Current patterns in the intrinsic frame for two rotational states
  in the three considered nuclei. For each nucleus.  The same arbitrary units
  for the length of $|\vec j|$ are used for both spins.  In all ellipses, the
  proton (resp. neutron) current patterns are represented in the top-right
  (resp. bottom-left) quarter. The global rotation direction is
  counter-clockwise. The formalisms used are HF, HF+V and HFB from top to
  bottom. The Lipkin-Nogami correction has been applied for $^{150}$Gd and
  $^{192}$Hg.}
  \label{fig:currents}
  \end{center}
\end{figure*}

As said in the Introduction, the Kelvin circulation is a powerful tool to
measure the intrinsic vortical content of the current distribution. In
Fig.~\ref{fig:currents} ($^{150}$Gd bottom part) we present the current
patterns in the intrinsic frame of two rotational states of the HFB
superdeformed yrast band of $^{150}$Gd, namely for $I=32\;\hbar$ and
$I=46\;\hbar$ (corresponding to angular velocities $\Omega=381$~keV and
$\Omega=516$~keV).
In both cases, the neutron current patterns are rather disordered (due to shell
structure effects). This is related to positive values of the neutron
contribution to the Kelvin circulations. On the contrary, proton current
patterns are rather well oriented, at least around the nuclear surface, along
elliptic lines parallel to the surface, the more so in the $I=32\;\hbar$ case
where total Kelvin circulation and pairing energies are greater (in absolute
value) than for $I=46\;\hbar$. This trend is due to the fact
(observed in Fig.~\ref{fig:kelvin}) that the proton HF contribution to
$|\langle\hat K_1\rangle|$ is very small.
In the HF case (upper part), the current patterns show no particular order for
both spins, and for protons as well as for neutrons. This clearly shows that
the inclusion of pairing correlations favors the appearance of
S-ellipsoid--like currents in rotating nuclei.

We then have performed HF+V calculations, namely constraining HF solutions to
have $I$ and $\langle\hat{K}_1\rangle$ values [see eqs. (\ref{eq:kelvin}) and
(\ref{eq:angular})] identical to their HFB counterparts. The constraint in use
is of an isoscalar type, that is constraining on the total (neutron + proton)
angular momentum and Kelvin circulation.
However, it can be seen in Fig.~\ref{fig:kelvin} that the proton and neutron
contributions to the Kelvin circulation in HF+V calculations are very close to
what is obtained in HFB calculations even though these contributions have not
been constrained separately. 
Moreover, the current patterns for the HFB and HF+V approaches (see
Fig.~\ref{fig:currents}) are similar. Hence we have successfully 
grafted the superfluid part of the HFB currents into the HF calculations
constraining merely the expectation values of the $\hat{K}_1$ isoscalar
operator.

The angular velocities within the three studied approaches, namely HF, HFB and
HF+V, are plotted in Fig.~\ref{fig:omega} against the angular momentum. It is
seen that the HFB and HF+V values fully overlap:
\begin{equation}
 \Omega _{HFB}\simeq \Omega _{HF+V}\,.
\label{eq:sameomega}
\end{equation}
It is worth noting that these two angular velocities are greater than
$\Omega_{HF}$. This is not surprising, since the counter-rotating collective
effect of pairing has to be compensated in HFB calculations at a given angular
momentum by an increase of the angular velocity. However, it is striking
to see that the identity of the two momenta of the current-distribution,
namely $I$ and $\langle\hat K_1\rangle$ in HFB and HF+V calculations, leads to
identical rotational angular velocities $\Omega$. It proves indeed, more
quantitatively than the plots of Fig.~\ref{fig:currents}, that the current
patterns in the two approaches are similar everywhere, which could hardly be
demonstrated by merely looking at integrated quantities.

The identity of the HFB and HF+V angular velocities seen in
Fig.~\ref{fig:omega} yields an identity of the kinetic moments
of inertia clearly demonstrated in eq. (\ref{eq:kinetic}). As for the HF+V
dynamical moments of inertia, even though they do not fully coincide with their
HFB counterparts, they are indeed very close and even reproduce better the
experimental data. In this nuclei, the HF+V and HFB formalisms yield solutions
whose axial quadrupole moments remain constant over the entire band (differing
by less than 2 percent around $Q_{20} \simeq 3900$~fm$^2$). Hence the
equivalence of the dynamical moments of inertia in the two formalisms is
clearly free from any deformation effect and reveals similar superfluid
properties.

\subsection{The yrast superdeformed band of $^{192}$Hg}
\label{ssec:mercure}
We have used also the HF, HFB+LN and HF+V formalisms in order to describe
the yrast superdeformed band of $^{192}$Hg and compare their results
with the available experimental data \cite{Gall94:hg}.

As already noted by Gall {\it et al.} \cite{Gall94} the very low amount of
pairing correlations attained above $I\simeq30\;\hbar$ (as exemplified by the
pairing energies for both neutrons and protons in the corresponding lower part
of Fig.~\ref{fig:kelvin}) makes it necessary to correct for the particle
number symmetry breaking.
This has been performed upon using the approximate Lipkin-Nogami particle
number projection on top of the HFB calculation which has been found necessary
to approach a proper description of the rotational band of this nuclei and
avoid spurious drops in the moments of inertia due to sudden pairing-energy
disappearance both for neutrons and protons.
Whereas the LN prescription is known to be efficient in restoring some pairing
correlations in weak-pairing regime as would be the case in HFB calculation
for these nuclear states, the question of knowing whether or not it provides
the correct amount of pairing correlations is still controversial (see, e.g.,
the discussion of Peru \cite{Per97:th}). A more exact treatment of pairing
correlations could be obtained for instance by generalizing to time-reversal
symmetry breaking systems the method of Pillet {\it et al.}
\cite{PQL02:htda,Pil00:th} which explicitly conserves the particle number.

\begin{figure*}[ht]
  \begin{center}
  \epsfig{width=17cm,file=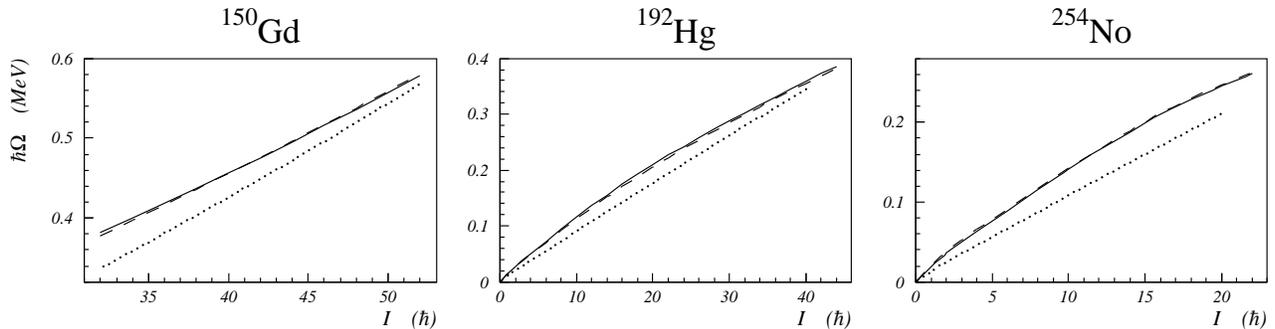}
  \caption{Angular velocity $\Omega$ (in MeV) as a function of the nuclear
  spin $I$ (in $\hbar$ units) for the three considered nuclei.  The
  conventions in use are the following: HF value (dotted line), HFB value
  (full line), and HF+V value (dashed line). The Lipkin-Nogami correction has
  been applied for $^{150}$Gd and $^{192}$Hg.}
  \label{fig:omega}
  \end{center}
\end{figure*}

The $^{192}$Hg dynamical moments of inertia are plotted in the upper panel of
Fig.~\ref{fig:inertia} within the three mentioned formalisms together with
their experimental counterparts.
As was already the case for $^{150}$Gd, the HF calculations fail to reproduce
the experimental values as well as the behavior of this moment as a function
of the angular velocity since it remains almost constant over the whole
rotational band.
By adding pairing correlations within the HFB+LN formalism, it is possible to
reproduce the experimental values at low spins. However, using a simple
seniority force as done here, one can only reproduce the increasing trend of
the experimental moment of inertia and our theoretical value becomes to high
at the upper bound of the rotational band.
In view of some residual instabilities of the solution due to the presence of
single-particle states in the vicinity of the cutoff energy defining the
configuration space where pairing correlations are treated, we have introduced
for this nucleus a cutoff factor of the Fermi-function type.

Similarly to what we have done for the $^{150}$Gd nucleus, we have attempted
to model the collective effects of pairing correlations in this nucleus by
performing HF+V calculations. As seen in Fig.~\ref{fig:inertia}, the
calculated HF+V dynamical moments of inertia nicely match the HFB+LN ones,
even better than for $^{150}$Gd. The axial deformation of the $^{192}$Hg
nuclear states in the rotational band calculated within these two formalisms
differ by less than 3 percent with an axial quadrupole moment mean value of
$Q_{20}\simeq 44$~barns.

In view of the rather low expectation values of the Kelvin circulation in the
HF regime (see Fig.~\ref{fig:kelvin}) its behavior as a function of the angular
momentum in the HFB approach reflects mostly the interplay between the global
rotation and the pairing correlations. One observes indeed a very interesting
parabolic pattern. At low spins its low value is due to the fact that the
intensity of the intrinsic currents is obviously related to the intensity of
the global rotational currents which they are counteracting.
On the other hand these intrinsic currents are, as we have assumed and
demonstrated here, related with the intensity of pairing correlations. This
balance explains the appearance of a maximum for $|\langle\hat K_1\rangle|$ at
some ``intermediate'' spin value (around $I\simeq20\;\hbar$). From the plotted
value of $\langle\hat K_1\rangle$ for both charge states in
Fig.~\ref{fig:kelvin}, one may infer that as it was the case for $^{150}$Gd,
the pairing generated intrinsic currents are almost entirely of an isoscalar
character.

The agreement between HFB and HF+V dynamical moments of inertia clearly points
into the direction of identical current patterns for both theoretical
descriptions. This is indeed so, as shown in Fig.~\ref{fig:currents}.
Finally, one observes in Fig.~\ref{fig:omega} that the similarity of the
kinetic moments of inertia in the two theoretical descriptions is reflected in
the corresponding curves giving the spin-dependence of the angular velocity.

\subsection{The ground-state deformed band of $^{254}$No}
\label{ssec:nobel}
Recently, the existence of the rotational ground-state band of the heavy
nucleus $^{254}$No has been assessed \cite{Rei99:no,*Lei99:no,*Rei00:no} by an
observation of the $\gamma$-transition energies from spins $I=4\;\hbar$ up to
$I=20\;\hbar$ in coincidence with the $\alpha$-decay chain of this
nucleus. This high spin structure is the first one studied in $Z>100$ nuclei.

It has been shown by several authors
\cite{Laf01:nobel,Dug01:nobel,Egid00:nobel} with different effective forces
and pairing parameterizations that the properties of this isotope could be
reproduced within HFB-type formalisms. The calculations presented here have
been performed in the spirit of Ref. \cite{Laf01:nobel}, that is fixing the
pairing-force parameterization in order to reproduce the first transition
energies. As a result, it happens that higher transition energies are fairly
well reproduced.  This procedure has been found in \cite{Laf01:nobel} to be
more successful within a pure HFB than a HFB+LN formalism. Thus we will
discuss here in terms of intrinsic vortical currents the results of our HFB
calculations.

The dynamical moments of inertia obtained within the three formalisms HF, HFB
and HF+V are plotted in the corresponding upper panel of
Fig.~\ref{fig:inertia} as functions of the angular velocity. As was the case
for the two previously studied nuclei, it is still impossible here to
reproduce the experimental data with pure HF calculations. On the contrary,
the results of our HFB calculations fit rather well in the experimental
patterns, and the HF+V results fully overlap the HFB ones. A slightly less
good agreement between both theoretical estimates of ${\frak J}^{(1)}$ among
themselves and with their experimental counterpart is obtained as exhibited in
Fig.~\ref{fig:inertia}.

In Fig. \ref{fig:kelvin}, the HF and HFB estimates of $\langle\hat K_1\rangle$
for the $^{254}$No neutron current distributions are seen to be very
close. This is clearly related as exhibited in the lower part of this figure
to the very small amount of pairing correlations for this charge state at spin
less than $6\;\hbar$ (whereas for higher spin values, there are no correlated
solutions).  As for the protons now, the shell effect contribution to
$\langle\hat K_1\rangle$ (i.e., the HF value) is quite significant, actually
of the same order of magnitude as the pairing contribution (i.e., the
difference between the HFB and the HF value).
As a very specific feature of the HF plus vorticity calculations in this
nucleus (versus those above reported in $^{150}$Gd and $^{192}$Hg) one notices
that a purely isoscalar constraint on $\langle\hat K_1\rangle$ is not
effective in reproducing the HFB results. Actually, whereas it is tempting to
attribute such an isovector behavior for a $Z\sim 100$ nucleus (as opposed to
$Z\sim 80$ or less nuclei) to the increasing relative importance of the
Coulomb interaction (scaled as $Z^2$), the observed behavior of
$\delta|\langle\hat K_1\rangle|$ (difference between HFB and HF+V values of
$|\langle\hat K_1\rangle|$) which is positive for protons and negative for
neutrons is yet to be understood (and should be assessed by systematic
calculations).

In Fig.~\ref{fig:currents} intrinsic currents corresponding to two very
different rotational regimes ($I=2$ and $20\;\hbar$) are displayed for
$^{254}$No. At low spins the neutron currents are small yet roughly consistent
with a tangential pattern at the surface. This is due to the existence of a
low pairing correlations regime. At high spin of course the HF and HFB neutron
current patterns are coinciding and hardly tangential at the surface. This
situation is quite at variance with what is obtained for the protons at large
spins, where the prevalence of large pairing correlations entails the
existence of rather well marked tangential counter-rotating currents.

Finally, the behavior of the angular velocity of both HFB and HF+V
calculations on the one hand and of the HF calculations on the other hand, as a
function of the total angular momentum is found quite similar here (see
Fig.~\ref{fig:omega}) to what was found before for $^{150}$Gd and $^{192}$Hg.

\section{Conclusions}
\label{sec:conclu}
In the above, we have examined in a Routhian approach, the equivalence of the
dynamical effects yielded by pairing correlations {\it \`a la} Bogoliubov
with those obtained when a specific ansatz for the collective flow is imposed
onto a purely Hartree-Fock type of microscopic calculations.
This has been achieved in rather different contexts as far as the nucleon
number ($A$ ranging from 150 to 250), the deformation (considering normally
deformed bands as well as superdeformed bands), the range of spin values as
well as the pairing correlation contents.
In all cases however, we have found rather convincingly that constraining
Hartree-Fock Routhian solutions to have the currents obtained by HFB Routhian
calculations at the same spin values, was a sound substitute to these HFB
calculations themselves.

For that purpose, the so-called Kelvin circulation operator has appeared as an
efficient measure of the intrinsic currents. Since this operator is well
suited to the collective dynamics dubbed as S-type ellipsoids by Chandrasekhar
(implying a collective field which is linear in the coordinates), the actual
HFB dynamical pattern should not deviate very much from this simple model
pattern.
That has been actually assessed by a mere inspection of current lines obtained
in our HFB calculations in cases where at large enough spin values strong
pairing correlations where present. Pairing correlations appear indeed as
generating tangential counter-rotating currents in the intrinsic frame.
This equivalence has been actually confirmed by a convergent array of
evidences:
\begin{enumerate}
\item[i)] the similarity of dynamical moments of inertia,
\item[ii)] the congruence of kinetic moment of inertia, or equivalently (at
the same value of the angular momentum) of the associated angular velocity,
\item[iii)] the correspondence between currents obtained in both approaches.
\end{enumerate}

As for the currents, it has appeared that shell effects where visible in
situations where either the collective intrinsic flow was too small (at low
angular velocity for the global rotation) or when pairing correlations where
vanishing.
These shell effects generating density inhomogeneities are responsible for
erratic current patterns (sometimes roughly similar to localized vortices)
which are in general neither counter-rotating nor tangential.

As above discussed, upon increasing the spin value one experiences a
competition between two processes.
The first one is an increase of the intrinsic currents, a reaction effect
which should be (and actually is, all things kept equal) increasing with the
global rotation angular velocity. This type of currents decreases, as well
known, the kinetic energy at a given angular velocity.
The second one is the Mottelson-Valatin
rotation-generated reduction of the pairing correlations on.  The competition
arises from the fact that this intrinsic flow should be decreasing with the
decrease of the pairing correlations.
One may thus easily understand the maximum obtained for the absolute value of
the expectation value of the Kelvin circulation obtained for the $^{192}$Hg
nucleus at intermediate spin values.

In our calculations an isoscalar constraint on the Kelvin circulation operator
has been used. Whereas it has been deemed to be adequate for the two lighter
nuclei under study, this has been hardly the case for the heavier
one. Confirming and, if so, understanding this $A$-dependence or more likely
$Z$-dependence, would require a rather systematic study which has
not been performed here.

The present study is now extended in two directions.

At high spins the decrease of pairing correlations has often been extrapolated
towards the appearance of a phase transition (associated thus with a complete
disappearance of these correlations). It is clearly not very appropriate to
quantitatively assess this feature in traditional approaches because one
reaches there a low pairing regime where the Bogoliubov particle number
symmetry-breaking approach is not very adequate there as well known.
We are currently undertaking \cite{Rhtda} Routhian calculations which
explicitly conserve the particle number according to the general lines
proposed in Ref. \cite{PQL02:htda}.

Even though we are representing here in a rather simple way the dynamical
effects of pairing correlations at finite angular momentum as obtained in a
Bogoliubov approach, we are still needing such lengthy calculations to yield
the correct amount of intrinsic currents (through the expectation value of the
Kelvin circulation operator).
The challenge is to try to correlate for a given nucleus (i.e., for
given nucleonic numbers, given deformations and given single particle level
densities near the chemical potential at zero spin or equivalently the amount
of pairing correlations at zero spin) the Kelvin circulation expectation value
as a function of the angular velocity of the global rotation.
This is currently achieved through a simple model approach inspired from the
Mottelson-Valatin rotation anti-pairing effect and will be presented soon in
an other paper \cite{LSQM2}.

\section{ACKNOWLEDGMENTS}
Part of this work has been funded through an agreement (\# 12533) between the
BAS (Bulgaria) and the CNRS (France) and another (\# 97-30) between the JINR
(Russia) and the IN2P3/CNRS (France) which are gratefully acknowledged. One
author (P. Q.) would like to thank the Theoretical Division of the LANL for the
hospitality extended to him during his stay at Los Alamos.
Two of the authors (H. L. and P. Q.) also wish to thank the
INT (Seatle)
for their kind invitation and the opportunity provided there to discuss some
parts of this paper.

\newdimen\rshift
\settowidth{\rshift}{\hbox to \columnwidth{ }\hbox to\columnsep{ }}
\end{document}